\documentclass[pdftex,twocolumn,epjc3]{svjour3}          

\RequirePackage[T1]{fontenc}

\smartqed  

\RequirePackage{graphicx}
\RequirePackage{mathptmx}      
\RequirePackage{flushend}
\RequirePackage[numbers,sort&compress]{natbib}
\RequirePackage[colorlinks,citecolor=blue,urlcolor=blue,linkcolor=blue]{hyperref}
\usepackage{graphicx}
\usepackage{epsfig}
\usepackage{amssymb}
\usepackage{amsmath}
\usepackage{subfigure}
\usepackage{epstopdf}
\usepackage{graphics}
\usepackage{epstopdf}
\usepackage{txfonts}
\usepackage{float}

\journalname{Eur. Phys. J. C}

\begin{document}

\title{Holographic Van der Waals phase transition of the higher dimensional electrically charged hairy black hole}

\subtitle{}

\author{Hui-Ling Li\thanksref{addr1,e1}
        \and
        Zhong-Wen Feng\thanksref{addr2}
}

\thankstext{e1}{e-mail: LHL51759@126.com}

\institute{College of Physics Science and Technology,
Shenyang Normal University, Shenyang 110034, China \label{addr1}
\and College of Physics and Space Science,
China West Normal University, Nanchong 637002, China \label{addr2}
}

\date{Received: date / Accepted: date}

\maketitle

\begin{abstract}
With motivation by holography, employing black hole entropy, two point connection function and entanglement entropy, we show that, for the higher dimensional Anti-de Sitter charged hairy black hole in the fixed charged ensemble, a Van der Waals-like phase transition can be observed. Furthermore, based on the Maxwell's equal area construction, we check numerically the equal area law for a first order phase transition in order to further characterize the Van der Waals-like phase transition.
\end{abstract}

\section{Introduction}

According to the Anti-de Sitter space/Conformal Field Theory (AdS/CFT) correspondence \cite{r1}, an asymptotically AdS black hole plays a key role in comprehending black hole thermodynamics of holographic dual field theories. One of the compelling phenomena of thermodynamics is phase transitions in AdS spacetime. The pioneer discussion on phase transition was done by Hawking and Page \cite{r2}. In 1983, in five dimensional non-charged AdS background, Hawking and Page proved that, there existed a first order phase transition between the AdS and Schwarzschild-AdS black hole. Specifically speaking, the thermal AdS is unstable, and it has to undergo a phase transition to the stable Schwarzschild-AdS black hole at last, which is the best known Hawking-Page phase transition.

Later, in 1990, in charged AdS background, Chamblin et al. \cite{r3,r4} explored the phase structure of an Reissner-Nordstr\"{o}m-AdS black hole. Compared to the non-charged AdS case, the charged black hole's phase structure becomes richer and is associated with the chosen statistical ensemble. It was found that the charged AdS black hole in the entropy-temperature plane presented an analogous Van der Waals phase transition in the canonical ensemble (fixed electric charge), i.e. the isocharge in the temperature-entropy plane has an unstable branch and two stable ones when the charge below a critical value. And there exists a second-order critical point at a critical charge. Recently, the research on a Van der Waals-like phase transition has been generalized to the extended phase space \cite{r5,r6,r7,r8,r9,r10}. In this framework, the cosmological constant is taken as a thermodynamical pressure, and its conjugate quantity is treated as the thermodynamical volume. Consequently, this extended phase space makes the Van de Waals description more precise.

Very recently, entanglement entropy has also been used to detect phase structures in different AdS backgrounds. Johnson \cite{r11} proposed that, like black hole entropy, entanglement entropy also exhibited a Van der Waals-like phase transition in the temperature-entanglement entropy plane in both the fixed potential ensemble and charge ensemble. It was subsequently generalized to supergravity STU black holes that involve four charges by Caceres et al. \cite{r12}. The result showed that, Van der Waals behavior was observed in the cases of three charge and four charge, however, for the one-charge and two-charge cases, the STU black holes did not present this phase transition. Caceres et al. also verified that for charge configuration that presented a Van der Waals-like phase transition, the entanglement entropy indeed exhibited a similar phase transition at the same critical temperature and the same critical exponents as the ones obtained in the black hole entropy. Furthermore, the equal area law of entanglement entropy was also checked by Nguyen \cite{r13} for AdS-Reissner-Nordstr\"{o}m black hole in the canonical ensemble. In addition, Zeng et al. have investigated phase structures of holographic entanglement entropy in massive gravity \cite{r14}, in the Born-Infeld-Anti-de Sitter background \cite{r15} and in the quintessence Reissner-Nordstr\"{o}m-AdS background \cite{r16}, and all the results showed that there existed a Van de Waals-like phase transition in these gravity backgrounds \cite{r14,r15,r16,Zeng:2017zlm,Zeng:2016aly}.

In the framework of holography, it is interesting to detect the phase structure of a higher dimensional charged AdS black hole. In the present work, we will focus on investigating the holographic phase transition of a five dimensional charged hairy black hole. Since there exist hair parameters, hairy black hole solutions become far richer than in General Relativity \cite{r17,r18}. Here, employing black hole entropy, two point correlated function and entanglement entropy, we attempt to study whether the Van der Waals-like phase transition can be observed, and discuss on the hair parameter's effect on phase transition.

\section{Phase transition and  Maxwell's equal area law for the thermodynamic entropy}

Let us start by reviewing the exact solution of an electrically charged hairy black hole in five dimensions. Einstein-Maxwell-$\Lambda$ theory conformally coupled to a scalar field in higher dimensions can exhibit analytic solutions \cite{r19,r20}. Here, we are concerned with the case of five dimensions, and the action of the theory reads

\begin{eqnarray} \label{1}
I=\frac{1}{\kappa}\int{\rm d}^{\,5}x\sqrt{-g}\left[R-2\Lambda-\frac{1}{4} F^{2}+\kappa L_{m}\left(\phi,\nabla\phi\right)\right],
\end{eqnarray}
where
\begin{eqnarray} \label{2-4}
&&\kappa=16 \pi G,\\
&&L_{m}\left(\phi,\nabla\phi\right)=b_{0}\phi^{15}+b_{1}\phi^{7} S_{\mu\nu}{}^{\mu\nu}+b_{2}\phi^{-1}\big(S_{\mu\gamma}{}^{\mu\gamma}S_{\nu\delta}{}^{\nu\delta}{}\nonumber\\{}&&\qquad\qquad\quad\,\,\,
-4S_{\mu\gamma}{}^{\nu\gamma}S_{\nu\delta}{}^{\mu\delta}+S_{\mu\nu}{}^{\gamma\delta}S^{\nu\mu}{}_{\gamma\delta}\big),\\
&&S_{\mu\nu}{}^{\gamma\delta}=\phi^{2}R_{\mu\nu}{}^{\gamma\delta}-12\delta{}^{[\gamma}_{[\mu}\delta_{\,\nu]}^{\delta]}
\nabla_{\rho}\phi\nabla^{\rho}\phi{}\nonumber\\{}&&\qquad\quad\,\,\,\,\,-48\phi\delta{}^{[\gamma}_{[\mu}\nabla_{\nu]}
\nabla^{\delta]}\phi+18\delta{}^{[\gamma}_{[\mu}\nabla_{\nu]}\phi\nabla^{\delta]}\phi.
\end{eqnarray}
Here, $b_{0}$, $b_{1}$ and $b_{2}$ are real coupling constants of the conformal field theory. The static spherically symmetric black hole solution coming from the action (1) can be written as
\begin{eqnarray} \label{5}
{\rm d}s^{\,2}=-N^{2}\left(r\right)f\left(r\right){\rm d}t^{\,2}+\frac{{\rm d}r^{\,2}}{g\left(r\right)}+r^{\,2}{\rm d}\Omega^{2}_{3},
\end{eqnarray}
where
\begin{eqnarray} \label{6}
g\left(r\right)=f\left(r\right)=1-\frac{m}{r^{\,2}}-\frac{q}{r^{\,3}}+\frac{e^{\,2}}{r^{\,4}}+\frac{r^{\,2}}{l^{\,2}},~~~~~~~~~N^{2}\left(r\right)=1.
\end{eqnarray}
Here, $l$ is AdS radius $l^{2}=-6/\Lambda$, and ${\rm d}\Omega^{2}_{3}$ is the metric of the unit 3-sphere.The integration constants $m$ and $e$ are related the mass and the electric charge of the hairy black hole, and $q$ is given with respect to the scalar coupling constants by the relation
\begin{eqnarray} \label{7}
q=\frac{64 \pi G}{5}\varepsilon b_{1}\left(-\frac{18 b_{1}}{5 b_{0}}\right)^{3/2},
\end{eqnarray}
where $\varepsilon=-1,~~0,~~+1$. For the five dimensional black hole solution to exist, the scalar coupling constants must obey the following constraint
\begin{eqnarray} \label{8}
 10 b_{0}b_{2}=9 b^{2}_{1}.
\end{eqnarray}
The Maxwell potential is
\begin{eqnarray} \label{9}
A_{\mu}=\sqrt{3}\frac{e}{r^{\,2}} \delta^{\,0}_{\,\mu}
\end{eqnarray}
with $F_{\mu\nu}=\partial_{\mu} A_{\nu}-\partial_{\nu} A_{\mu}$. On the other hand, the scalar field configuration takes the form
\begin{eqnarray} \label{10}
\phi\left(r\right)=\frac{n}{r^{1/3}},~~~~~~n=\varepsilon\left(-\frac{18}{5}\frac{b_{1}}{b_{0}}\right)^{1/6}.
\end{eqnarray}
In Ref. \cite{r19}, Galante et. al have discussed on the thermodynamic properties of higher dimensional black holes in detail. The mass and charge of the hairy black hole are
\begin{eqnarray} \label{11}
M=\frac{3\pi}{8} m=\frac{3\pi\left(e^{2}l^{2}-q l^{2}r_{+}+l^{2}r^{4}_{+}-r^{6}_{+}\right)}{8l^{2}r^{2}_{+}}
\end{eqnarray}
and
\begin{eqnarray} \label{12}
Q=-\frac{\sqrt{3}\pi}{8}e,
\end{eqnarray}
respectively, where $r_{+}$ is the event horizon, which is given by the equation $f\left(r_{+}\right)=0$. The Hawking temperature and black hole entropy are given by \cite{r7,r19}
\begin{eqnarray} \label{13-14}
&&T=\frac{1}{\pi l^{2}r^{4}_{+}}\left(-\frac{32Q^{2}l^{2}}{3\pi^{2}r_{+}}+\frac{q l^{2}}{4}+\frac{l^{2}}{2}r^{3}_{+}+r^{5}_{+}\right),\\
&&S=\frac{\pi^{2}}{2}\left(r^{3}_{+}-\frac{5}{2}q\right).
\end{eqnarray}
Note that, for the hairy black hole solution to exist, $q$ can take three different values namely $q=0,~~\pm \left|q\right|$. For $ q=0$, the solution reduces to the five dimensional Reissner-Nordstr\"{o}m AdS black hole, and the phase transition has been discussed on extensively \cite{r13,r21}. Consequently, here we mainly focus on studying the phase structure for the case $q\neq0$.

Now, we begin to explore the hairy black hole's critical behavior and phase transition in the temperature--entropy plane. From Eqs. (13) and (14), we can get the function $T(S,\,Q,\,q)$ by eliminating $r_{+}$
\begin{eqnarray} \label{15}
T(S,Q,q)=\!\!\!\!&&\frac{1}{3\sqrt[3]{2}l^{2}\pi^{5/3}(5\pi^{2}q+4S)^{5/3}}\Big[75\pi^{4}q^{2}{}\nonumber\\{}&&-128l^{2}\pi^{2}Q^{2}+120\pi^{2}Sq+48S^{2}
{}\nonumber\\{}&&+9\sqrt[3]{4}l^{2}\pi^{10/3}q(5\pi^{2}q+4S)^{1/3}{}\nonumber\\{}&&+6\sqrt[3]{4}l^{2}\pi^{4/3}S(5\pi^{2}q+4S)^{1/3}\Big].
\end{eqnarray}
Note that the phase structure of a hairy black hole is not only related to electric charge $Q$, but also the scalar hair parameter $q$. Based on the function $T(S,\,Q,\,q)$ above, the phase structure of the charged hairy black hole can be detected. We find out that, by choosing some proper hair parameters $q$, the isocharges in the $T-S$ plane can exhibit a Van der Waals-like phase transition. For convenience, we keep the AdS radius $l=1$ throughout this paper. Here, we discuss on the phase transition in the fixed electric charged ensemble, and take $q=\pm 0.010$ and $q=\pm 0.005$ as examples. To plot the isocharges in the $T-S$ plane, firstly we should get the critical values of phase transition by using the following relations
\begin{eqnarray} \label{16}
\left(\frac{\partial T}{\partial S}\right)_{Q}=\left(\frac{\partial^{2} T}{\partial^{2} S}\right)_{Q}=0.
\end{eqnarray}
However, it is hard to obtain analytical values directly. We have to get them numerically. In Table1, we tabulate the critical charge $Q_{C}$, critical entropy $S_{C}$ and critical temperature $T_{C}$ for different hair parameter $q$. From this table, we see that the critical entropy becomes smaller as $q$ increases. According to Eq. (14), we know that, $q$ must satisfy $q\leqslant 2r\,_{+}^{3}/5$ in order for the positivity condition of entropy.

According to the relation in (15) and these critical values, we plot the isocharges in the $T-S$ plane for different $q$ in Fig. 1. Each curve corresponds to a different electric charge. Remarkably, as can be seen from these plots, a Van der Waals-like phase transition is clearly presented in the $T-S$ plane. For different $q$, the phase structure is similar. when $Q>Q_{C}$, temperature is monotonically larger as entropy increases, and the system is thermodynamically stable. As $Q$ decreases and arrive to the critical Value $Q_{C}$, an inflection point appears and the heat capacity is divergent at this point, which corresponds to an second order phase transition. In the $Q<Q_{C}$ case, in addition to two stable branches, the isocharge has an unstable branch with negative heat capacity
\begin{eqnarray} \label{17}
C_{Q}=T\left(\frac{\partial S}{\partial T}\right)_{Q},
\end{eqnarray}
which corresponds to a first order phase transition.

\begin{figure}

\centering
\subfigure[$q=-0.010$]{
\begin{minipage}[b]{0.38\textwidth}
\includegraphics[width=1\textwidth]{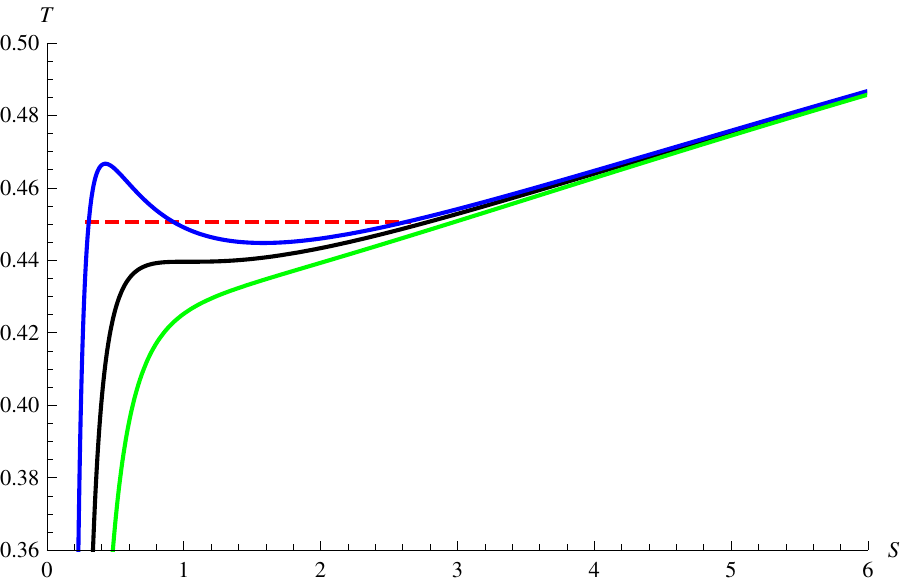}
\end{minipage}
}\hspace{6ex}
\subfigure[$q=-0.005$]{
\begin{minipage}[b]{0.38\textwidth}
\includegraphics[width=1\textwidth]{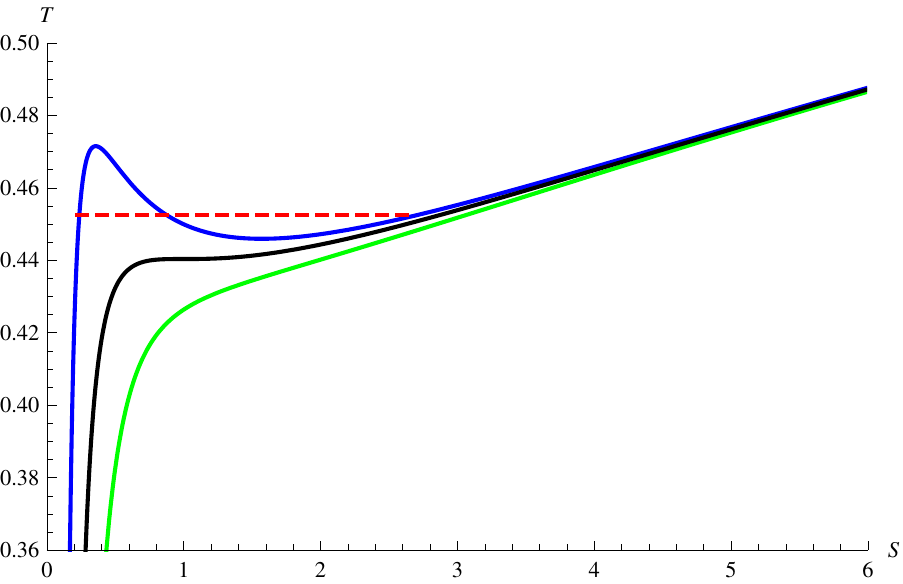}
\end{minipage}
}\vfill
\subfigure[$q=0.005$]{
\begin{minipage}[b]{0.38\textwidth}
\includegraphics[width=1\textwidth]{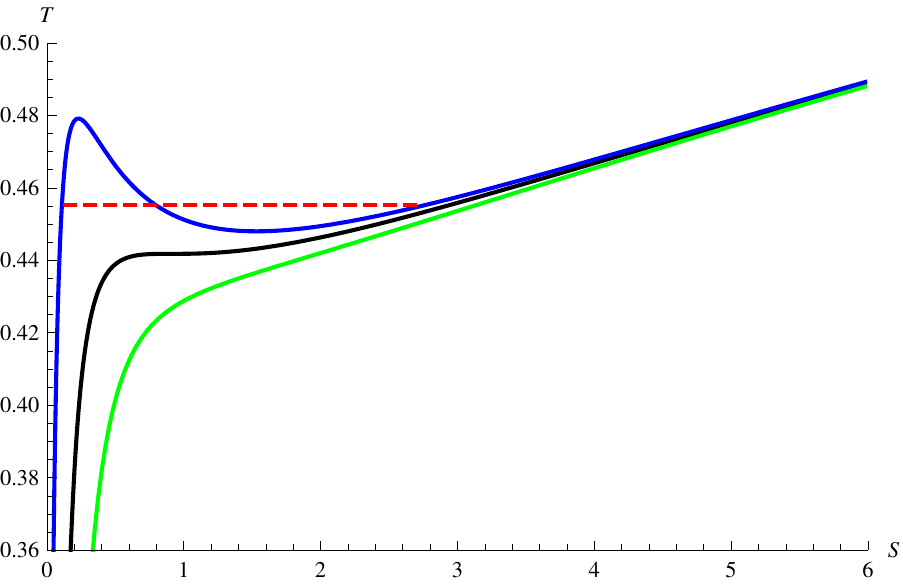}
\end{minipage}
}\hspace{7ex}
\subfigure[$q=0.010$]{
\begin{minipage}[b]{0.38\textwidth}
\includegraphics[width=1\textwidth]{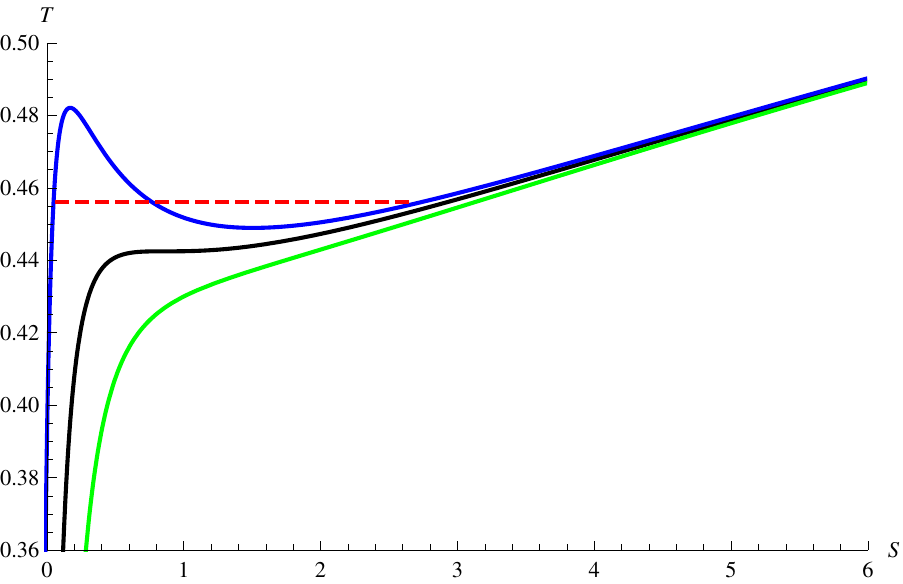}
\end{minipage}
}

\caption{Plots of the temperature versus the black hole entropy for different q. The red dash line corresponds to the temperature of the first order phase transition. The values of the electric charge chosen (from top to bottom) are as follows.
Panel (a): $Q=0.0286384<Q_{C}$, $Q=0.0486384=Q_{C}$, $Q=0.0686384>Q_{C}$. Panel (b): $Q=0.0338006<Q_{C}$, $Q=0.0538006=Q_{C}$, $Q=0.0738006>Q_{C}$. Panel (c): $Q=0.0429513<Q_{C}$, $Q=0.0629513=Q_{C}$, $Q=0.0829513>Q_{C}$. Panel (d): $Q=0.0470977<Q_{C}$, $Q=0.0670977=Q_{C}$, $Q=0.0870977>Q_{C}$}
\label{fig1}
\end{figure}

\begin{figure}

\centering
\subfigure[$q=-0.010$]{
\begin{minipage}[b]{0.4\textwidth}
\includegraphics[width=1\textwidth]{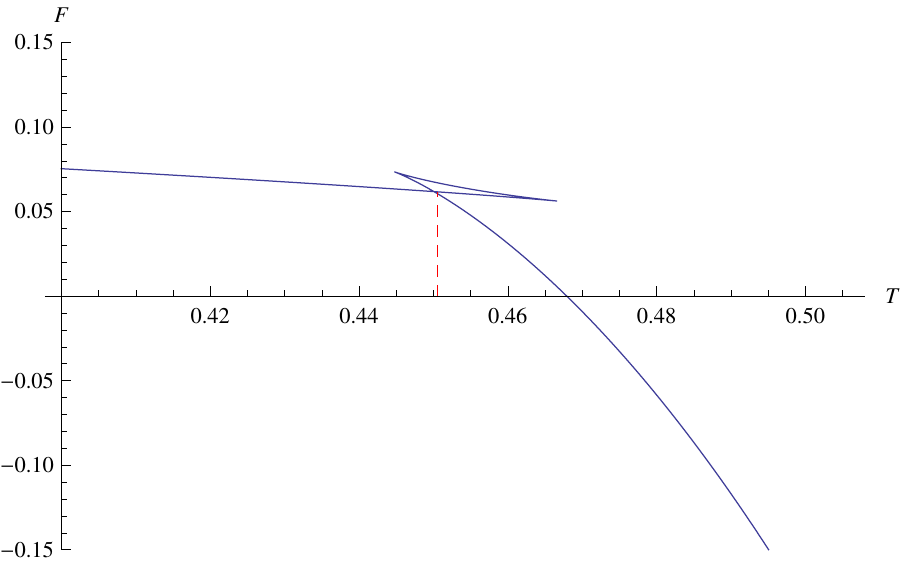}
\end{minipage}
}\hspace{6ex}
\subfigure[$q=-0.005$]{
\begin{minipage}[b]{0.4\textwidth}

\includegraphics[width=1\textwidth]{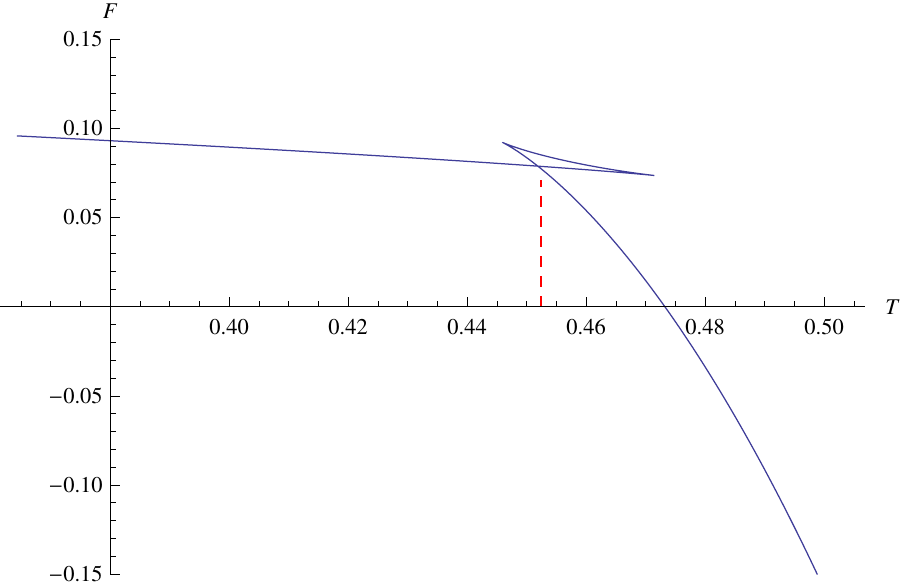}
\end{minipage}
}\vfill
\subfigure[$q=0.005$]{
\begin{minipage}[b]{0.4\textwidth}
\includegraphics[width=1\textwidth]{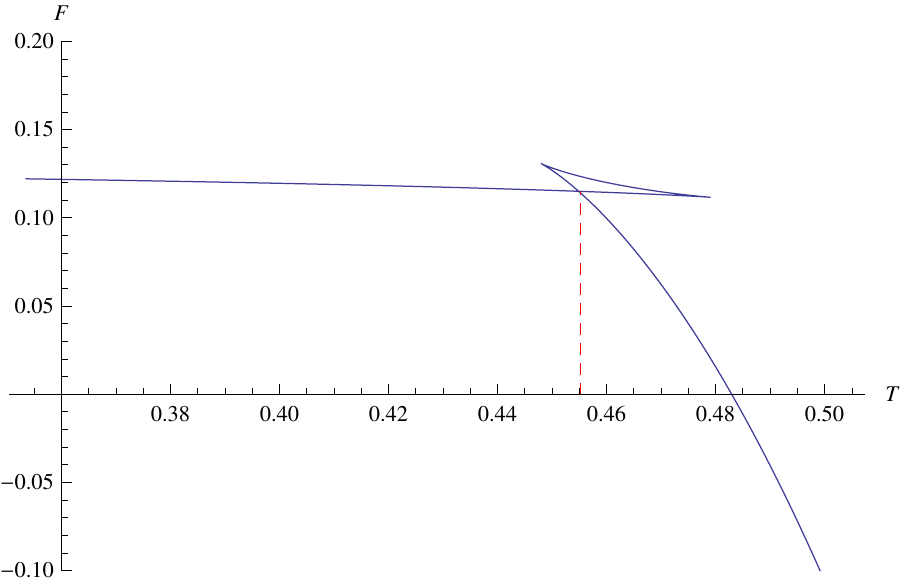}
\end{minipage}
}\hspace{7ex}
\subfigure[$q=0.010$]{
\begin{minipage}[b]{0.4\textwidth}
\includegraphics[width=1\textwidth]{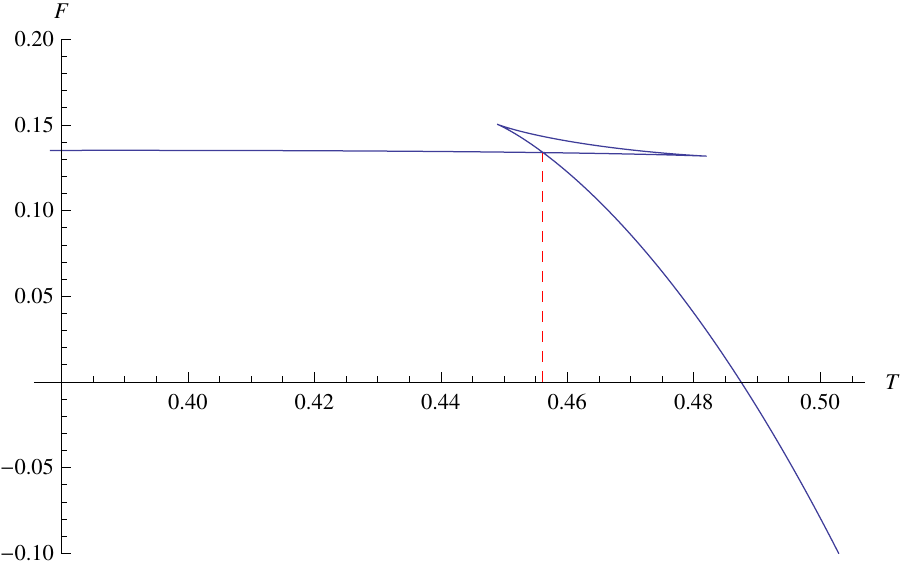}
\end{minipage}
}

\caption{Plots of the temperature versus the free energy for different q in case of $Q<Q_{C}$. In each graph, the red dash line indicates a first order phase transition temperature $T^{*}$. Panel (a): $Q=0.0286384$ and $T^{*}=0.4506$. Panel (b): $Q=0.0338006$ and $T^{*}=0.4524$. Panel (c): $Q=0.0429513$ and $T^{*}=0.4552$. Panel (d): $Q=0.0470977$ and $T^{*}=0.4561$}
\label{fig2}
\end{figure}

\begin{table}
\centering
\caption{The critical values of the electric charge, temperature and entropy for different $q$}
\label{parset}
\begin{tabular*}{\columnwidth}{@{\extracolsep{\fill}}llll@{}}
\hline
\multicolumn{1}{c}{$q$} & \multicolumn{1}{c}{$Q_{C}$} & \multicolumn{1}{c}{$T_{C}$} & \multicolumn{1}{c}{$S_{C}$} \\
\hline
$-$0.010           & 0.0486384  & 0.439591       & 1.03476         \\
$-$0.005             & 0.0538006     & 0.440337        & 0.992571         \\
\,\,\,\,$0.005$           & 0.0629513  & 0.441770       & 0.906233         \\
\,\,\,\,$0.010$             & 0.0670977        & 0.442460      & 0.862219        \\
\hline
\end{tabular*}
\end{table}

Like the Van der Waals phase transition of the liquid-gas system, this unstable portion should be replaced by using an isotherm $T=T^{*}$ which obeys Maxwell's equal-area prescription. The subcritical temperature $T^{*}$ can be obtained from the plot of the free energy $F=M-TS$ versus the temperature. The plots in Fig. 2 show that the relations between the temperature and free energy for different $q$, and for $Q<Q_{C}$, we always observe a classic swallowtail structure in each plot, which is responsible for the first order phase transition in Fig. 1. We indicate the transition temperature $T^{*}$ by a red dashed line in Fig. 2, which corresponds to the horizontal coordinate of the junction.

Next, in order to further characterize the Van der Waals-like phase transition, we turn to check Maxwell equal area law for the first order phase transition and the corresponding statement can be written as
\begin{eqnarray} \label{18}
A_{1}\equiv \int _{S_{min}}^{S_{max}}T(S,\,Q,\,q)\,{\rm d}S=T^{*}(S_{max}-S_{min})\equiv A_{2},
\end{eqnarray}
where $T(S,\,Q,\,q)$ is defined in Eq. (15), $S_{min}$ and $S_{max}$ are the smallest and largest roots of the equation $T(S,\,Q,\,q)=T^{*}$. Now, we take $q=-0.010$ as an example to show how to verify the Maxwell's equal area law. Numerically, using $q=-0.010$, $Q=0.0286384$ and $l=1$, we obtain $T^{*}$ from the plot (a) in Fig. 2. Then, substituting $T^{*}=0.4506$ into Eq. (15), we obtain the smallest value $S_{min}=0.306264$ and largest value $S_{max}=2.628909$ by resolving the equation $T(S,-0.01,0.0286384)=0.4506$. Thus,using the values of $T^{*}$, $S_{min}$ and $S_{max}$, we can get $A_{2}=1.04658$ at right side of Eq. (18), and obtain  $A_{1}=1.04538$ by integrating left side of Eq. (18). Namely, $A_{1}$ equals to $A_{2}$ in our numeric accuracy with these values. Repeating the procedure above, we can obtain $A_{1}$ and $A_{2}$ for other $q$, and we tabulate these values in Table 2. From this table, it is obvious that the Maxwell's equal area construction holds in the $T-S$ plane.

\section{Phase transition and Maxwell's equal area law for two point correlation function}
In this section, we proceed to discuss on the phase structure of two point correlation function. In recent years, the two point correlation function appears to be a useful tool which can be used to explore some physical phenomena such as holographic singularities \cite{r22,r23}, holographic thermalization \cite{r24,r25}, holographic CFTs on maximally symmetric spaces \cite{r26}, holographic butterfly effect
\cite{Shenker:2013pqa,Leichenauer:2014nxa, Sircar:2016old,Cai:2017ihd} and quantum phase transition \cite{r27}. Motivated by the mentioned above, we attempt to detect whether the two point correlation function can present a Van der Waals-like behaviour for the charged hairy black hole.

According to the Anti-de Sitter space/Conformal Field Theory dictionary, in the large $\triangle$ limit, the equal time two point correlation function can be holographically expressed as \cite{r28}
\begin{eqnarray} \label{19}
\left\langle O(t_{0},x_{i})\,O(t_{0},x_{j})\right\rangle\approx e^{-\triangle L},
\end{eqnarray}
where $\triangle$ is the conformal dimension of scalar operator $O$, and $L$ is the length of the buck geodesic between the point $(t_{0},\,x_{i})$ and $(t_{0},\,x_{j})$ on the Anti-de Sitter boundary. Due to the charged hairy black hole's symmetry in five dimensional AdS spacetime, we can simply set $(\theta=\theta_{\,0},\,\varphi=\pi/2,\,\psi=0)$ and $(\theta=\theta_{\,0},\,\varphi=\pi/2,\,\psi=\pi)$ as the two boundary point.

By utilizing $\theta$ to parameterize the trajectory, the proper length takes the form
\begin{eqnarray} \label{20}
L=\int_{0}^{\theta_{0}}\mathcal{L}(r(\theta),\theta)\,{\rm d}\theta, \quad \mathcal{L}=\sqrt{\frac{r\,'^{\,2}(\theta)}{g(r)}+r^{\,2}(\theta)},
\end{eqnarray}
here $r'\equiv {\rm d}r/{\rm d}\theta$.

\begin{figure}[H]

\centering
\subfigure[$q=-0.010$]{
\begin{minipage}[b]{0.4\textwidth}
\includegraphics[width=1\textwidth]{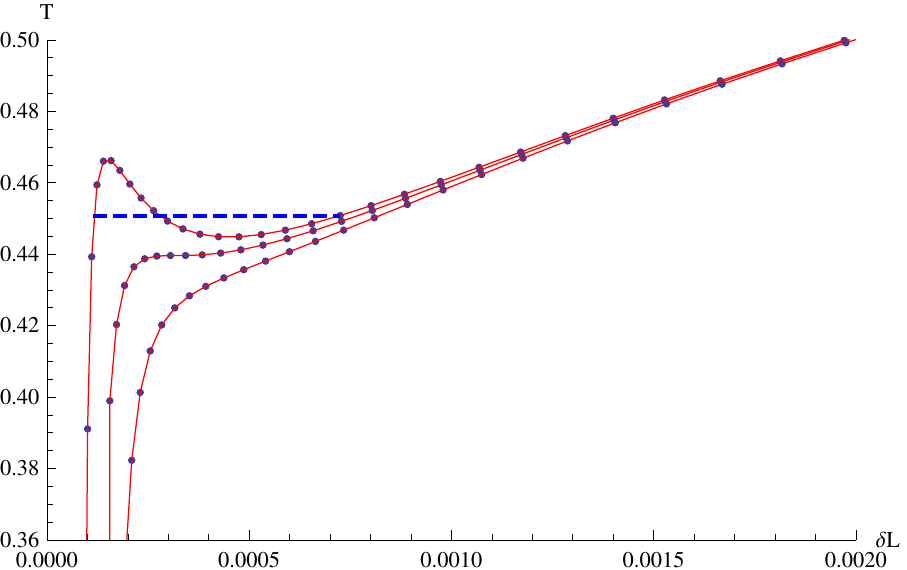}
\end{minipage}
}\hspace{6ex}
\subfigure[$q=-0.005$]{
\begin{minipage}[b]{0.4\textwidth}
\includegraphics[width=1\textwidth]{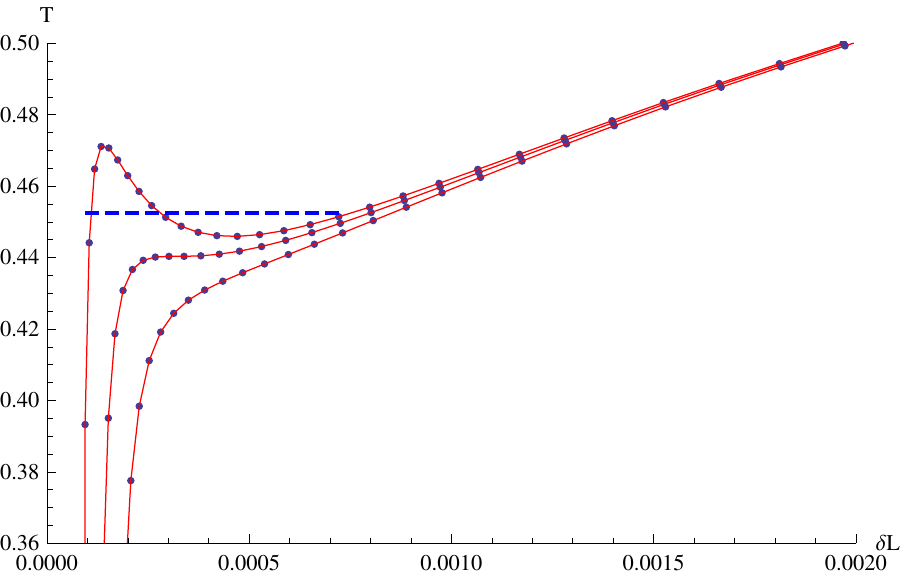}
\end{minipage}
}\vfill
\subfigure[$q=0.005$]{
\begin{minipage}[b]{0.4\textwidth}
\includegraphics[width=1\textwidth]{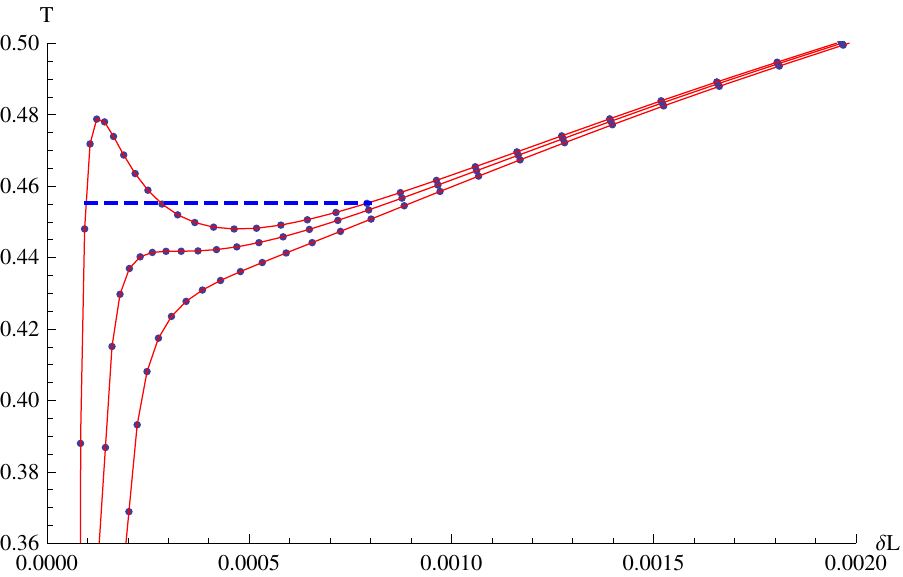}
\end{minipage}
}\hspace{7ex}
\subfigure[$q=0.010$]{
\begin{minipage}[b]{0.4\textwidth}
\includegraphics[width=1\textwidth]{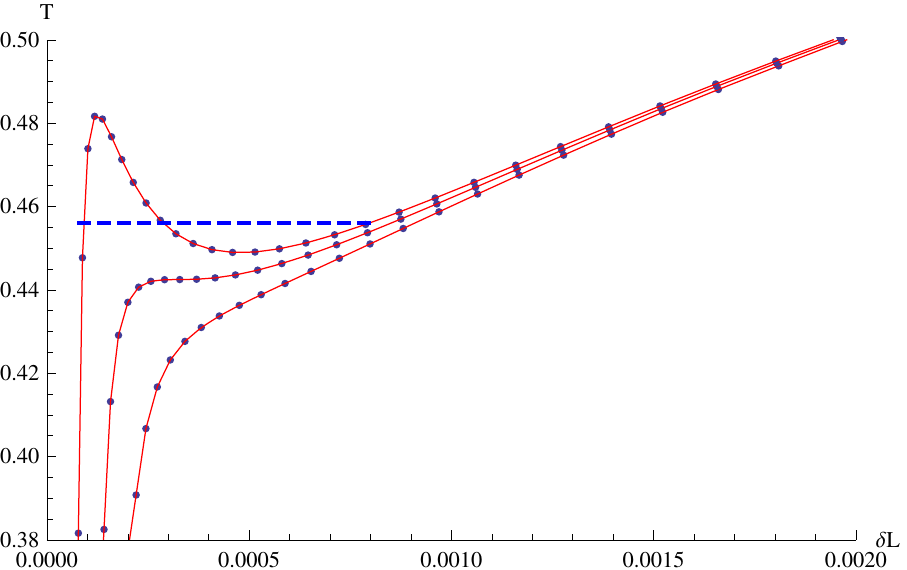}
\end{minipage}
}

\caption{Plots of isocharges in $T-\delta L$ plane for $\theta_{\,0}=0.45$. The Blue dash line corresponds to the temperature of the first order phase transition. The values of the electric charge chosen (from top to bottom)are the same with ones in Fig. 1}
\label{fig3}
\end{figure}

\begin{figure}[H]

\centering
\subfigure[$q=-0.010$]{
\begin{minipage}[b]{0.4\textwidth}
\includegraphics[width=1\textwidth]{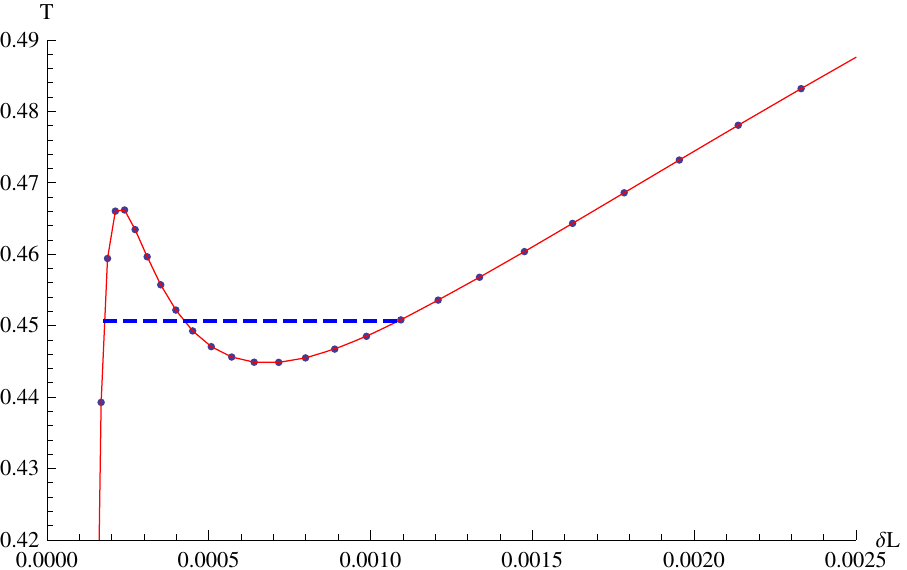}
\end{minipage}
}\hspace{6ex}
\subfigure[$q=-0.005$]{
\begin{minipage}[b]{0.4\textwidth}
\includegraphics[width=1\textwidth]{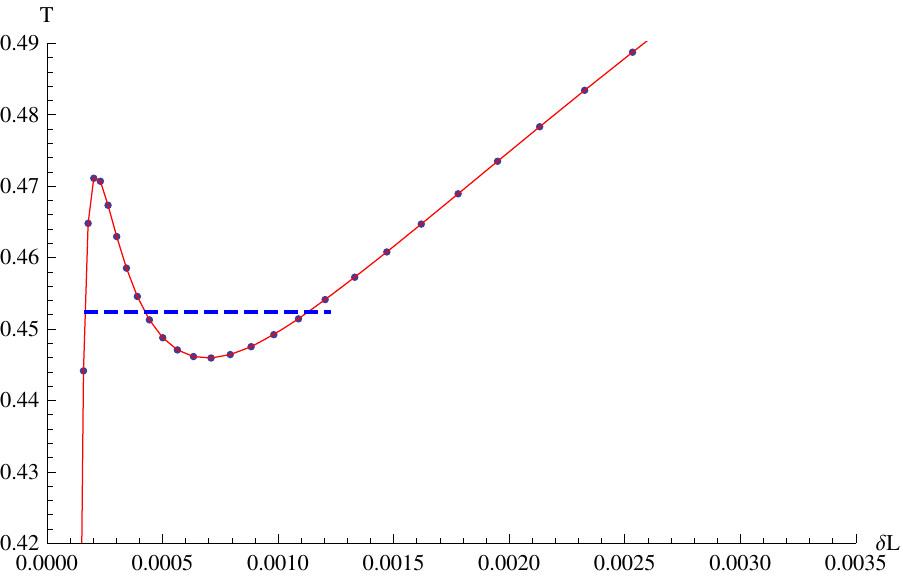}
\end{minipage}
}\vfill
\subfigure[$q=0.005$]{
\begin{minipage}[b]{0.4\textwidth}
\includegraphics[width=1\textwidth]{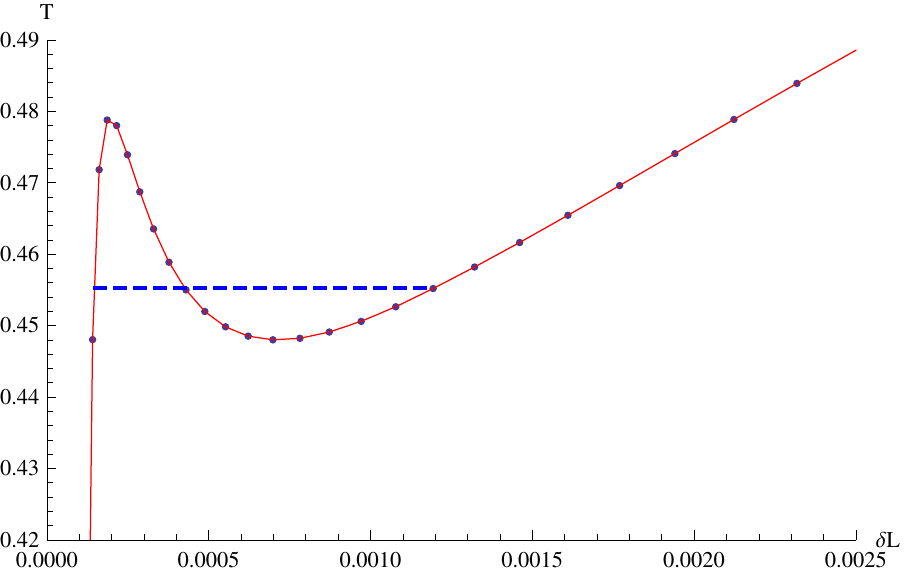}
\end{minipage}
}\hspace{7ex}
\subfigure[$q=0.010$]{
\begin{minipage}[b]{0.4\textwidth}
\includegraphics[width=1\textwidth]{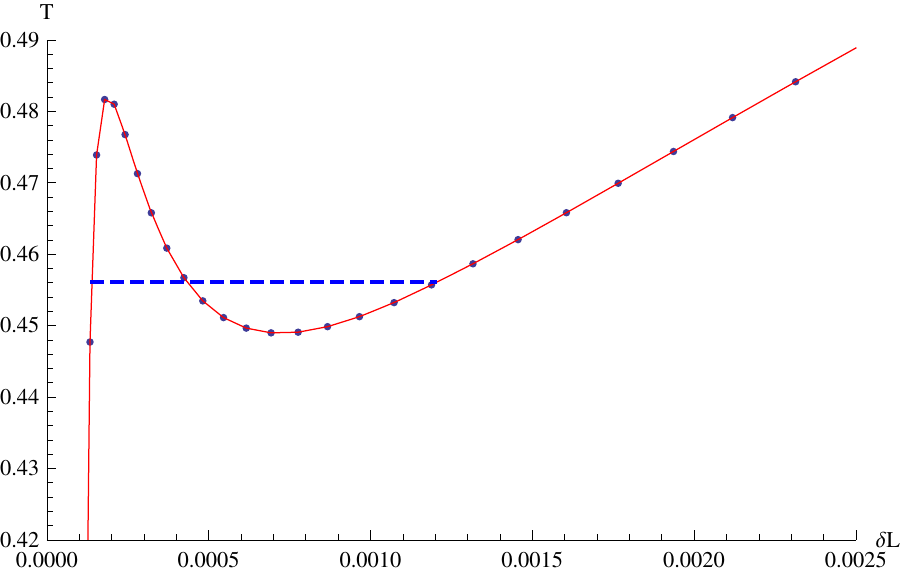}
\end{minipage}
}

\caption{Plots of isocharges in $T-\delta L$ plane for $\theta_{\,0}=0.50$. The Blue dash line corresponds to the temperature of the first order phase transition for $Q<Q_{C}$.  Panel (a): $Q=0.0286384<Q_{C}$.  Panel (b): $Q=0.0338006<Q_{C}$.  Panel (c): $Q=0.0429513<Q_{C}$.  Panel (d): $Q=0.0470977<Q_{C}$}
\label{fig4}
\end{figure}

\begin{figure}[H]

\centering
\subfigure[$q=-0.010$]{
\begin{minipage}[b]{0.4\textwidth}
\includegraphics[width=1\textwidth]{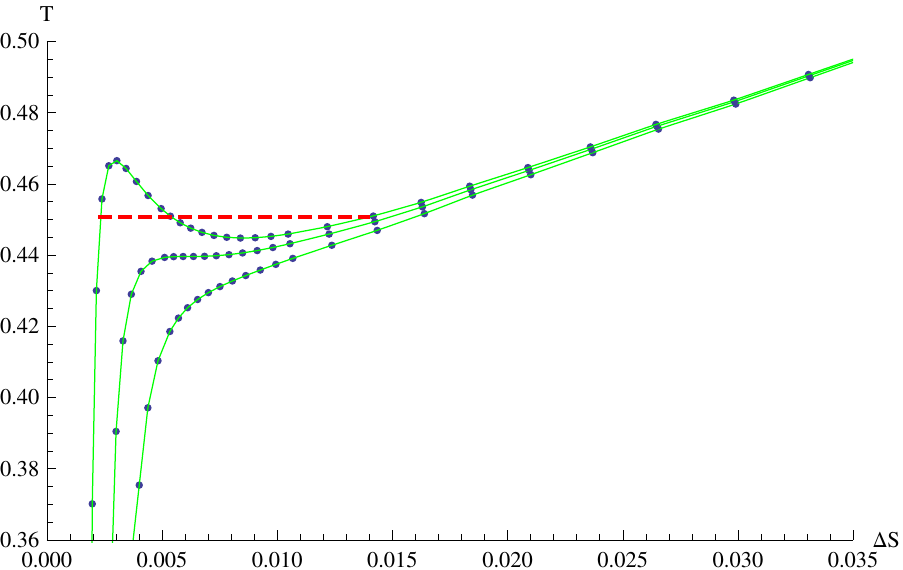}
\end{minipage}
}\hspace{6ex}
\subfigure[$q=-0.005$]{
\begin{minipage}[b]{0.4\textwidth}
\includegraphics[width=1\textwidth]{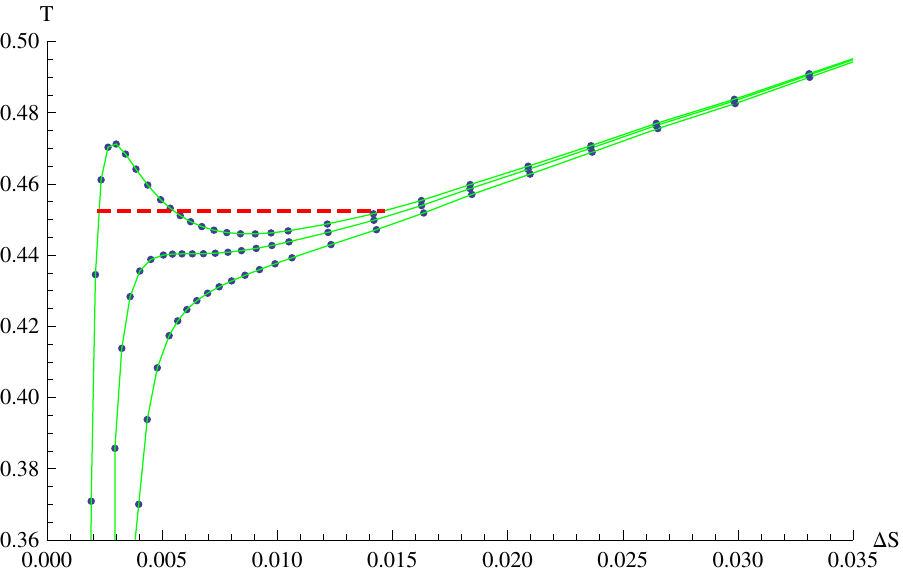}
\end{minipage}
}\vfill
\subfigure[$q=0.005$]{
\begin{minipage}[b]{0.4\textwidth}
\includegraphics[width=1\textwidth]{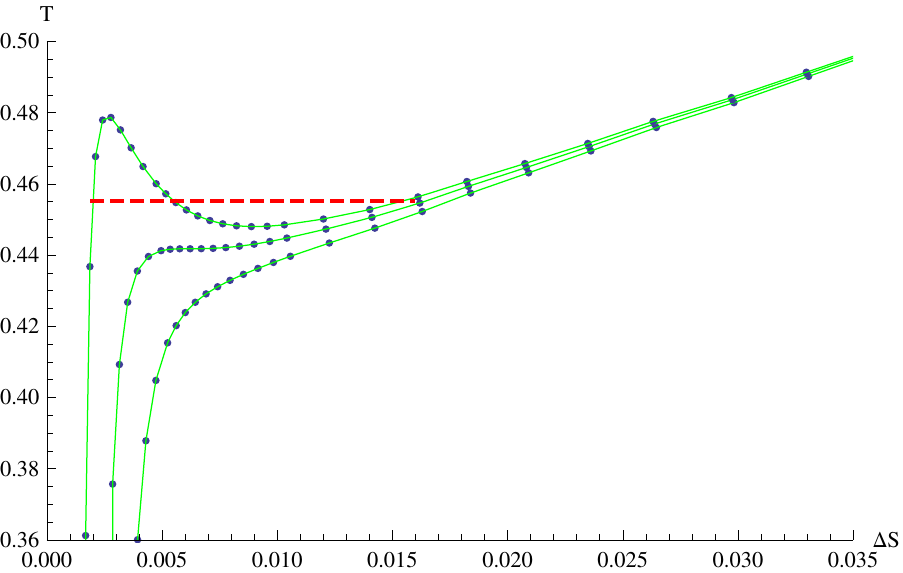}
\end{minipage}
}\hspace{7ex}
\subfigure[$q=0.010$]{
\begin{minipage}[b]{0.4\textwidth}
\includegraphics[width=1\textwidth]{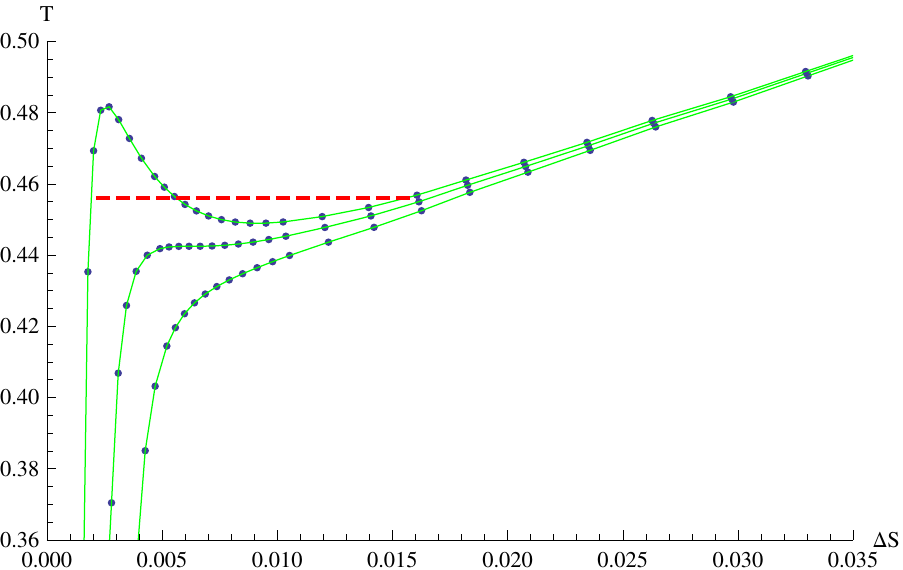}
\end{minipage}
}

\caption{Plots of isocharges in $T-\triangle S$ plane for $\varphi_{\,0}=0.45$. The red dash line corresponds to the temperature of the first order phase transition. The values of the electric charge chosen (from top to bottom) are the same with ones in Fig. 1}
\label{fig5}
\end{figure}

\begin{figure}[H]

\centering
\subfigure[$q=-0.010$]{
\begin{minipage}[b]{0.4\textwidth}
\includegraphics[width=1\textwidth]{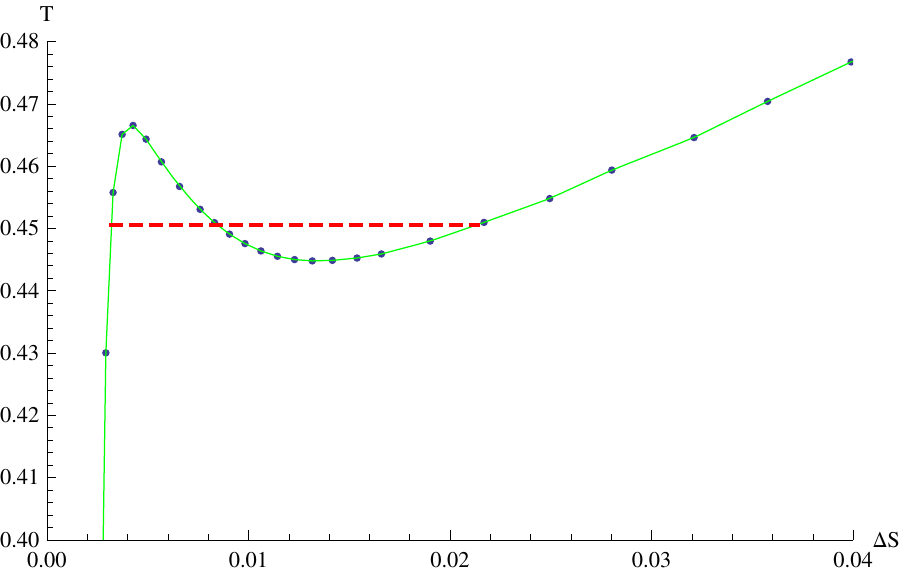}
\end{minipage}
}\hspace{6ex}
\subfigure[$q=-0.005$]{
\begin{minipage}[b]{0.4\textwidth}
\includegraphics[width=1\textwidth]{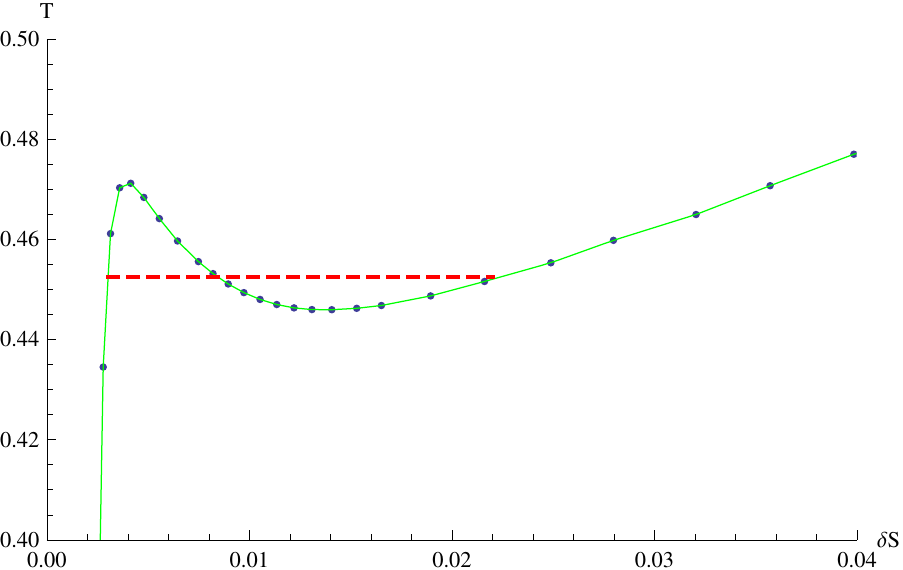}
\end{minipage}
}\vfill
\subfigure[$q=0.005$]{
\begin{minipage}[b]{0.4\textwidth}
\includegraphics[width=1\textwidth]{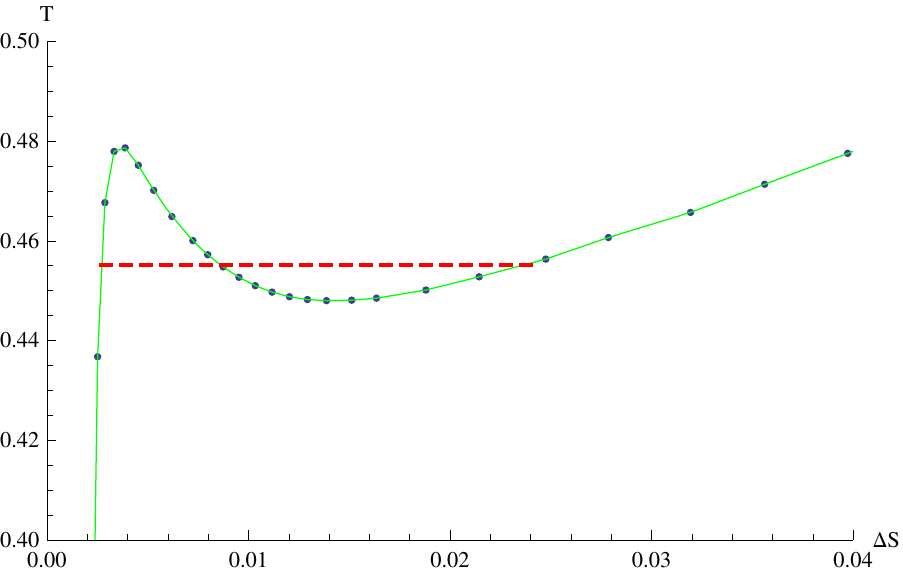}
\end{minipage}
}\hspace{7ex}
\subfigure[$q=0.010$]{
\begin{minipage}[b]{0.4\textwidth}
\includegraphics[width=1\textwidth]{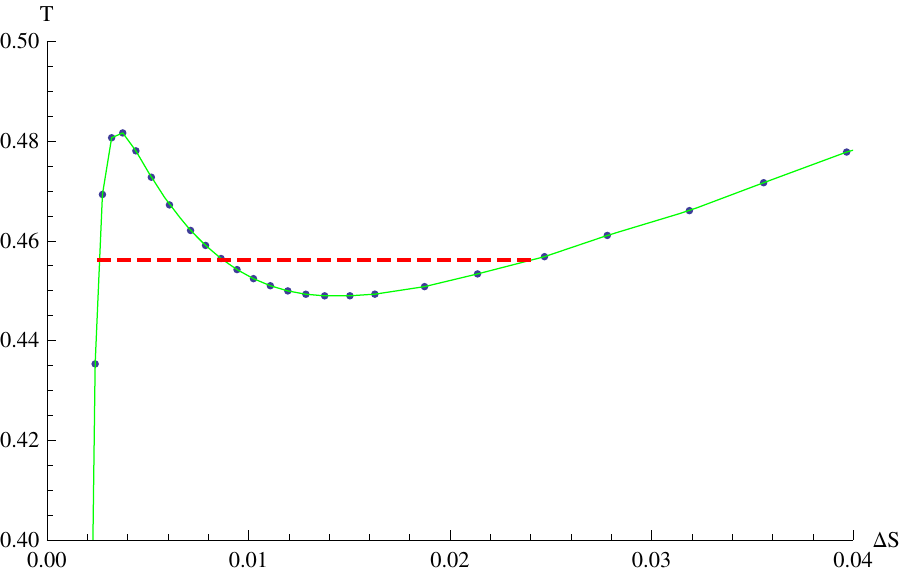}
\end{minipage}
}

\caption{Plots of isocharges in $T-\triangle S$ plane for $\varphi_{\,0}=0.50$. The red dash line corresponds to the temperature of the first order phase transition for $Q<Q_{C}$. Panel (a): $Q=0.0286384<Q_{C}$.  Panel (b): $Q=0.0338006<Q_{C}$.  Panel (c): $Q=0.0429513<Q_{C}$.  Panel (d): $Q=0.0470977<Q_{C}$}
\label{fig6}
\end{figure}

\begin{table*}[htbp]
\caption{Check of the Maxwell's equal area construction in the $T-S$ plane}
\label{sphericcase}
\begin{tabular*}{\textwidth}{@{\extracolsep{\fill}}lrrrrl@{}}
\hline
\multicolumn{1}{c}{$q$} & \multicolumn{1}{c}{$T^{*}$} & \multicolumn{1}{c}{$S_{min}$} & \multicolumn{1}{c}{$S_{max}$} & \multicolumn{1}{c}{$A_{1}$} & \multicolumn{1}{c}{$A_{2}$}  \\
\hline
$-$0.010 & 0.4506 & 0.3062640 & 2.6289090 & 1.04538 & 1.04658 \\
$-$0.005 &   0.4524   &  0.2371192 & 2.6989999 & 1.11250 & 1.11375 \\
\,\,\,\,0.005 &    0.4552 & 0.1098390 & 2.7624530 & 1.20665 & 1.20747 \\
\,\,\,\,0.010 &    0.4561 & 0.0488927 & 0.7663060 & 1.22943 & 1.22924 \\
\hline
\end{tabular*}
\end{table*}

Taking $\mathcal{L}$ as the Lagrangian and imagining $\theta$ as time, with the Euler-Lagrange equation
\begin{eqnarray} \label{21}
\frac{\partial \mathcal{L}}{\partial r}=\frac{{\rm d}}{{\rm d}\theta}\left(\frac{\!\!\!\!\!\!\!\!\!\!\!\!\!\partial \mathcal{L}}{\partial r\,'(\theta)}\right),
\end{eqnarray}
we arrive to the motion equation of $r(\theta)$
\begin{eqnarray} \label{22}
g\,'(r)r\,'(\theta)^{\,2}-2g(r)r''(\theta)+2g^{\,2}(r)r(\theta)=0
\end{eqnarray}
with the boundary condition
\begin{eqnarray} \label{23}
r(0)=r_{0}, \qquad r\,'(0)=0.
\end{eqnarray}
By using the condition above and resolving the Eq. (22), we obtain the mumeric result of $r(\theta)$. Notice that the geodesic length is divergent for a fixed $\theta_{0}$, therefore, it needs to be regularized. Here, we do it by subtracting the geodesic length of the minimal surface in pure AdS with same boundary $\theta=\theta_{0}$ (denoted by $L'$). In order to accomplish this, we first obtain $L$ by intergrating the length function in Eq. (20) from zero to $UV$ cutoff $\theta_{C}\lesssim\theta_{\,0}$. Then, by turning off the hair parameter $q$, mass $M$ and electric charge $Q$ of the charged hairy black hole background, we get pure AdS in global coordinates
\begin{eqnarray} \label{24}
{\rm d}s^{\,2}=-\left(1+\frac{r^{\,2}}{l^{\,2}}\right){\rm d}t^{2}+\left(1+\frac{r^{\,2}}{l^{\,2}}\right)^{-1}{\rm d}r^{\,2}+r^{\,2}{\rm d}\Omega_{3}^{2},
\end{eqnarray}
with this metric, repeating the same procedure as calculation $L$, numerically, we obtain $L'$. Thus, subtracting this quantity from the hairy black hole, we get the renormalized geodesic length $\delta L=L-L'$. Here, we take $\theta_{\,0}=0.45$ and $\theta_{\,0}=0.50$ as example to discuss the charged hairy black hole's phase structure, and the corresponding cutoffs are chosen to be $\theta_{C}=0.449$ and $0.499$ during the numerical computations.
For the $\theta_{\,0}=0.45$ case, in Fig. 3. we present the plots of the isocharges generated for the two point correlation function for different hair parameter $q$. In each panel, the isocharges in the $T-\delta L$ plane from top to bottom correspond to $Q<Q_{C}$, $Q=Q_{C}$ and $Q>Q_{C}$, respectively. From this plots, we see that the $T$ versus $\delta L$ plots are qualitatively similar to ones in Fig. 1. We confirm that there exists indeed a Van der Waals-like phase transition in the $T-\delta L$ plane, and the critical temperature and critical charge are also the same as the ones obtained from Fig. 1.

\begin{table*}
\caption{Check of the Maxwell's equal area construction in the $T-\delta L$ plane}
\label{sphericcase}
\begin{tabular*}{\textwidth}{@{\extracolsep{\fill}}lrrrrl@{}}
\hline
\multicolumn{1}{c}{$ $} & \multicolumn{1}{c}{$ $} & \multicolumn{1}{c}{$\theta_{\,0}=0.45$} & \multicolumn{1}{c}{$ $} & \multicolumn{1}{c}{$\theta_{\,0}=0.50$}   \\
\hline
$q=-0.010$ & $T^{*}=0.4506$ & $\delta L_{min}=0.0001181430$ & $A_{1}=0.00027069$ & $\delta L_{min}=0.0001788345$ & $A_{1}=0.00040808$ \\
$ $ &      & $\delta L_{max}=0.0007194292$  & $A_{2}=0.00027094$ & $\delta L_{max}=0.0010853080$ & $A_{2}=0.00040846$ \\
\hline
$q=-0.005$ & $T^{*}=0.4524$ & $\delta L_{min}=0.0001099755$ & $A_{1}=0.00028904$ & $\delta L_{min}=0.0001664403$ & $A_{1}=0.00043580$ \\
$ $ &      & $\delta L_{max}=0.0007494337$  & $A_{2}=0.00028929$ & $\delta L_{max}=0.0011305550$ & $A_{2}=0.00043617$ \\
\hline
$q=0.005$ & $T^{*}=0.4552$ & $\delta L_{min}=0.00009738140$ & $A_{1}=0.00031608$ & $\delta L_{min}=0.0001473907$ & $A_{1}=0.00047660$ \\
$ $ &      & $\delta L_{max}=0.00079196240$  & $A_{2}=0.00031617$ & $\delta L_{max}=0.0011947060$ & $A_{2}=0.00047674$ \\
\hline
$q=0.010$ & $T^{*}=0.4561$ & $\delta L_{min}=0.00009229728$ & $A_{1}=0.00032276$ & $\delta L_{min}=0.0001397911$ & $A_{1}=0.00048686$ \\
$ $ &      & $\delta L_{max}=0.00079951970$  & $A_{2}=0.00032256$ & $\delta L_{max}=0.0012065893$ & $A_{2}=0.00048657$ \\
\hline
\end{tabular*}
\end{table*}

\begin{table*}[htbp]
\caption{Check of the Maxwell's equal area construction in the $T-\triangle S$ plane}
\label{sphericcase}
\begin{tabular*}{\textwidth}{@{\extracolsep{\fill}}lrrrrl@{}}
\hline
\multicolumn{1}{c}{$ $} & \multicolumn{1}{c}{$ $} & \multicolumn{1}{c}{$\varphi_{\,0}=0.45$} & \multicolumn{1}{c}{$ $} & \multicolumn{1}{c}{$\varphi_{\,0}=0.50$}   \\
\hline
$q=-0.010$ & $T^{*}=0.4506$ & $\triangle S_{min}=0.002343243$ & $A_{1}=0.0052170$ & $\triangle S_{min}=0.00321284$ & $A_{1}=0.0081764$ \\
$ $ &      & $\triangle S_{max}=0.013932195$  & $A_{2}=0.0052220$ & $\triangle S_{max}=0.02136945$ & $A_{2}=0.0081814$ \\
\hline
$q=-0.005$ & $T^{*}=0.4524$ & $\triangle S_{min}=0.00227335$ & $A_{1}=0.0055859$ & $\triangle S_{min}=0.00302369$ & $A_{1}=0.0087256$ \\
$ $ &      & $\triangle S_{max}=0.01463187$  & $A_{2}=0.0055910$ & $\triangle S_{max}=0.02232081$ & $A_{2}=0.0087300$ \\
\hline
$q=0.005$ & $T^{*}=0.4552$ & $\triangle S_{min}=0.00201383$ & $A_{1}=0.0061112$ & $\triangle S_{min}=0.00273341$ & $A_{1}=0.0095445$ \\
$ $ &      & $\triangle S_{max}=0.01544459$  & $A_{2}=0.0061137$ & $\triangle S_{max}=0.02369990$ & $A_{2}=0.0095440$ \\
\hline
$q=0.010$ & $T^{*}=0.4561$ & $\triangle S_{min}=0.00193166$ & $A_{1}=0.0062515$ & $\triangle S_{min}=0.00261655$ & $A_{1}=0.0097623$ \\
$ $ &      & $\triangle S_{max}=0.01563169$  & $A_{2}=0.0062486$ & $\triangle S_{max}=0.02400015$ & $A_{2}=0.0097531$ \\
\hline
\end{tabular*}
\end{table*}

In order to further characterize a Van der Waals-like phase transition for the two point correlation function, we choose two different $\theta_{0}$ to verify  the equal area law in $T-\delta L$ plane, which is similarly defined as
\begin{eqnarray} \label{25}
A_{1}\equiv \int _{\delta L_{min}}^{\delta L_{max}}T(\delta L,\,q,\,Q)\,{\rm d}\delta L=T^{*}(\delta L_{max}-\delta L_{min})\equiv A_{2},\nonumber\\
\end{eqnarray}
where $\delta L_{min}$ and $\delta L_{max}$ are the smallest and largest roots of the equation $T(\delta L,\,q,\,Q)=T^{*}$. $T(\delta L,\,q,\,Q)$ is an interpolating function which can be given by our numeric result.

Besides $\theta_{\,0}=0.45$, in Fig. 4, for the case $\theta_{\,0}=0.50$, we also plot the isocharge in the $T-\delta L$ plane when the charge satisfies $Q<Q_{C}$. In table 3, for different $q$ and $\theta_{\,0}$, we tabulate the values of $\delta L_{min}$, $\delta L_{max}$, $A_{1}$ and $A_{2}$. Obviously, for the two point correlation function, the equal area law hods for the five dimensional charged hairy black hole. This strengthen our conclusion that, indeed, the isocharges in the $T-\delta L$ plane can also present the same Van der Waals-like phase transition as the black hole entropy.

\section{Holographic phase transition for entanglement entropy}
Now, we move on to take into account entanglement entropy case. According to Ryu-Takayanagi description, holographic entanglement entropy for a region $A$ can be expressed as \cite{r29,r30}
\begin{eqnarray} \label{26}
S=\frac{Area(\Gamma_{A})}{4},
\end{eqnarray}
where $\Gamma_{A}$ is a codimension-2 minimal surface with boundary condition $\partial\Gamma_{A}=\partial A$.
Here, we choose $\varphi=\varphi_{0}$ as the hairy black hole's entangling surface and employ $r(\varphi)$ to parameterize the minimal surface. With the symmetry, in this static AdS background, the entanglement entropy is given by
\begin{eqnarray} \label{27}
S=\pi\int_{0}^{\varphi_{\,0}}r^{\,2}\sin^{2}\varphi \sqrt{\frac{r\,'^{\,2}(\varphi)}{f(r)}+r^{\,2}(\varphi)}\,{\rm d}\varphi,
\end{eqnarray}
in which $r\,'(\varphi)\equiv {\rm d}r/{\rm d}\varphi$. Adopting the similar procedure as the two point correlation function case, we first arrive to the motion equation of $r(\varphi)$ by utilizing the Euler-Lagrange equation, and with the boundary condition, the numerical result of $r(\varphi)$ is obtained. Then, we integrate the entropy function $S$ in Eq. (27) up to the $UV$ cutoff $\varphi_{\,C}$ (that is $\varphi_{\,C}\approx \varphi_{\,0}$). Thus, substracting the pure AdS entanglement entropy (which is denoted by $S_{0}$), we are able to get the regularized entanglement entropy $\triangle S=S-S_{0}$ of a charged hairy black hole.

Specifically, choosing $\varphi_{\,0}=0.45$, we present the isocharges in the $T-\triangle S$ plane for different hair parameter $q$ in Fig. 5.
Comparing with Fig. 1 and Fig. 3 again, we find that, like the black hole entropy and two point correlation function, the entanglement entropy also exhibits a Van der Waals-like phase transition, moreover, the critical charge and critical temperature are also identified with them.

We go on to verify whether Maxwell construction works for the entanglement entropy in the $T-\triangle S$ plane. The analogous equal area law becomes
\begin{eqnarray} \label{28}
A_{1}\equiv \!\!\!\int _{\triangle S_{min}}^{\triangle S_{max}}\!\!\!T(\triangle S,q,Q)\,{\rm d}\triangle S\!=\!T^{*}(\triangle S_{max}\!-\!\triangle S_{min})\equiv A_{2},
\end{eqnarray}
where $\triangle S_{min}$ and $\triangle S_{max}$ are the smallest and largest roots of $T(\triangle S,\,q,\,Q)=T^{*}$, $T(\triangle S,\,q,\,Q)$ is an interpolating function which is given by our numeric result, and $T^{*}$ is a transition temperature, which is equal to the first order phase transition temperature $T^{*}$ found for black hole entropy in section 2. In order to verify the equal area law, we take $\varphi_{\,0}=0.45$ and $\varphi_{\,0}=0.50$ as example, we also present the isocharges in the $T-\triangle S$ plane for $\varphi_{\,0}=0.50$ in Fig. 6. For different $q$ and $\varphi_{0}$, the numeric results of $\triangle S_{min}$, $\triangle S_{max}$, $A_{1}$ and $A_{2}$ are listed in Table 4. According to this table, we conclude that the Maxwell's equal area construction in the $T-\triangle S$ plane is valid within reasonable error. Again, the result shows that like black hole entropy, the entanglement entropy can indeed present a Van der Waals-like phase transition for the charged hairy black hole.

\section{Conclusion}
In this paper, in the framework of holography, we discuss on the phase structure of a charged hairy black hole in five-dimensional AdS background (in the fixed electric charge ensemble). The result shows that a Van der Waals-like phase transition can be observed in $T-S$ plane, $T-\delta L$ plane and $T-\triangle S$ plane, and the critical charge and critical temperature are equal to each other. Notice that for some $q$ value, where the gravity background exhibits negative entropy, there no longer exists a reasonal phase transition. In order to guarantee the entropy is positive, scalar hair parameter $q$ must satisfy the condition $q\leqslant 2r\,_{+}^{3}/5$. Since the corresponding expression for the critical values are too complicated in this charged hairy AdS background, here, we proceed to compute numerically.

It is interesting to note that, for the charged hairy black hole, in Ref. \cite{r7}, Hennigar and Mann have first reveal a reentrant phase transition, and have carefully studied the criticality and the Van der Waals behaviour in the $P-V$ plane. Here, in the $T-S$ plane, by choosing some proper values of $q$, we also present the Van der Waals-like phase transition, thereby strengthening the conclusion of Ref. [7]. It is worth emphasizing that, motivated by holography, besides the black hole entropy, we also make use of the two point correlation function and entanglement entropy to detect the Van der Waals-like phase transition. In addition, we compute numerically the equal area law in the $T-S$ plane, $T-\delta L$ plane and $T-\triangle S$ plane, and verify the Maxwell's equal area construction is valid in these planes, which provides further evidence supporting our conclusion.

\begin{acknowledgements}
We would like to thank Xiao-Xiong Zeng for his discussions. This work is supported by the
National Natural Science Foundation of China (Grant Nos. 11573022).
\end{acknowledgements}

\end{document}